\documentclass[%
 reprint,
 amsmath,amssymb,
 aps,
superscriptaddress,]{revtex4-2}

\usepackage{graphicx}
\usepackage{dcolumn}
\usepackage{bm}
\usepackage{hyperref}
\usepackage{adjustbox} 
\usepackage{siunitx}
\usepackage{chemformula}
\usepackage{comment}

\begin{document}

\preprint{APS/123-QED}

\title{High-Stress Si$_3$N$_4$ Reflective Membranes\\ Monolithically Integrated with Cavity Bragg Mirrors}%

\author{Megha Khokhar}
\affiliation{Department of Precision and Microsystems Engineering, Delft University of Technology, Delft 2628 CD, The Netherlands}
\affiliation{Kavli Institute of Nanoscience, Department of Quantum Nanoscience, Delft University of Technology, Delft 2628 CD, The Netherlands}%
\author{Lucas Norder}
\affiliation{Department of Precision and Microsystems Engineering, Delft University of Technology, Delft 2628 CD, The Netherlands}
\author{Paolo M. Sberna} 
\email{p.m.sberna@tudelft.nl}
\affiliation{Else Kooi Laboratory, Faculty of Electrical Engineering, Mathematics and Computer Science, Delft University of Technology, Delft 2628 CD, The Netherlands}
\author{Richard A. Norte}
\email{r.a.Norte@tudelft.nl}
\affiliation{Department of Precision and Microsystems Engineering, Delft University of Technology, Delft 2628 CD, The Netherlands}

\date{\today}

\begin{abstract}

High-stress silicon nitride (Si$_3$N$_4$) membranes represent the state-of-the-art for cavity optomechanics, combining ultralow dissipation, optical transparency, and full compatibility with wafer-scale nanofabrication. Yet their integration into high-finesse optical cavities has remained difficult, typically requiring bonding or alignment-sensitive assembly that limits scalability and long-term stability. Here, we introduce a monolithic, wafer-level integration strategy that directly suspends high-stress Si$_3$N$_4$ photonic-crystal membranes above thermally compatible SiN/SiO$_2$ distributed Bragg reflectors (DBRs) capable of withstanding the high temperatures required for stoichiometric Si$_3$N$_4$ growth. A defect-free amorphous-silicon sacrificial layer and stiction-free plasma undercut yield vertically coupled cavities with sub-micron spacing—forming self-aligned resonators within seconds of release. Owing to the intrinsic tensile stress, the suspended membranes exhibit atomic-scale sagging, ensuring near-ideal cavity parallelism and long-term stability. Optical reflectivity measurements reveal cavity finesse exceeding $8\times10^2$ with nanoscale gaps between mirrors. Mechanical ringdown measurements show $Q_{\mathrm{mech}} > 10^{5}$, indicating that DBR integration preserves the low-dissipation character of high-stress Si$_3$N$_4$. This demonstrates that the integration process preserves the material’s exceptional dissipation dilution, supporting straightforward extension to high-Q nanomechanical architectures reported in the literature. The resulting Si$_3$N$_4$–DBR platform unites optical and mechanical coherence with high fabrication yield and design flexibility, enabling scalable optomechanical devices for precision sensing and quantum photonics.

\end{abstract}

\maketitle

\section{\label{sec:level1}Introduction}

\begin{figure*}[t]
\centering
\includegraphics[width=0.9\textwidth]{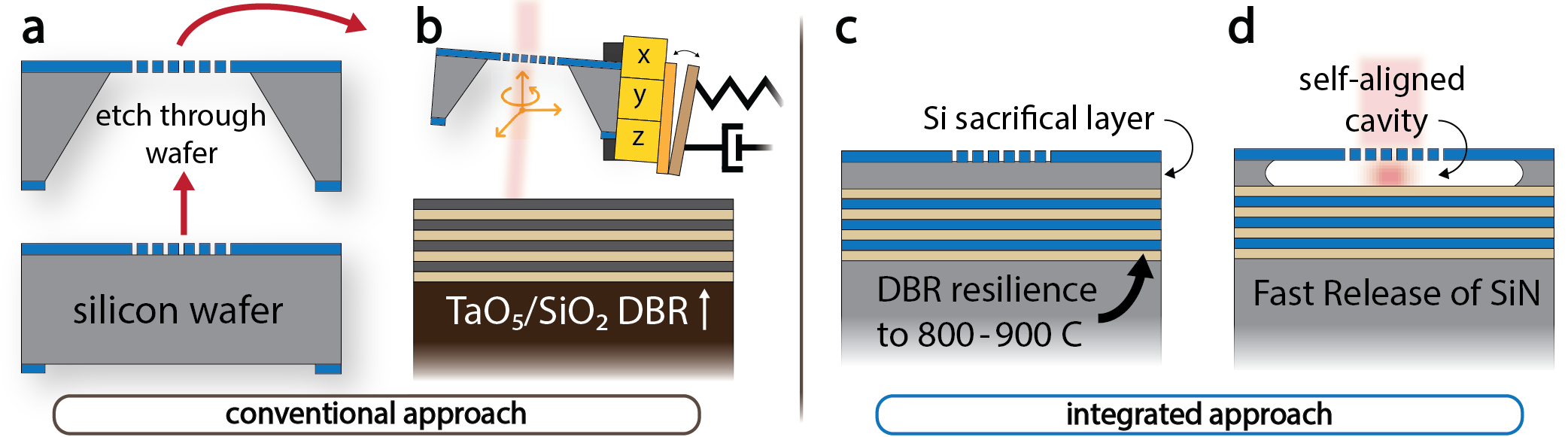}
\caption{Conventional fabrication of Si$_3$N$_4$ Membranes vs. integrated approach. (a) conventional fabrication of Si$_3$N$_4$ membranes for optical access requires deep etches through the chip which are complex, delicate and leave residues that must be subsequently cleaned off. (b) Complex alignment infrastructure to align with conventional DBR to form cavity. (c) integrated approach deposits high stress Si$_3$N$_4$ at high temperatures over Si sacrificial layer and DBR made from SiO$_2$ and Si$_3$N$_4$. A few second undercut of Si layer suspended high stress Si$_3$N$_4$ photonic crystal membranes above DBR to form self-aligned, monolithic optomechanical cavity with a high $Q_{mech}$.}
\label{fig0}
\end{figure*}

Cavity optomechanics, which leverages the interaction between light and mechanical motion in resonant systems, has emerged as a pivotal platform for both fundamental studies and precision sensing technologies~\cite{RevModPhys.86.1391,LiOuLei2021}. At its simplest, it consists of an optical cavity with one of its mirrors being movable. The mirror’s motion modulates the cavity field enabling exquisite readout of displacement and forces. Confining light within small volumes with high-finesse cavities further amplifies these effects, enabling ground-state cooling~\cite{rossi2018}, quantum squeezing~\cite{huang2024room}, and ultra-sensitive force~\cite{Reinhardt2016} and some of the most sensitive displacement detection achieved in science~\cite{abbott2016observation}. Achieving such performance requires extreme optical and mechanical coherence: the cavity must maintain high finesse, and the mechanical element must exhibit ultralow mechanical dissipation and strong optomechanical coupling. To approach these limits, the movable mirror should combine high optical quality (high reflectivity and low curvature) with low mechanical loss, which in practice favors lightweight, high-stress membranes that can be engineered for minimal coupling to their supports.   

Among available materials for such reflectors, high-stress silicon nitride (Si$_3$N$_4$) membranes have emerged as a state of the art for room-temperature optomechanics~\cite{engelsen2024ultrahigh}. Their combination of large built-in tensile stress~\cite{xu2024high}, ultralow optical absorption at telecom wavelengths (parts per billion)~\cite{land2024sub}, and wafer-scale CMOS manufacturing has enabled exceptional optomechanical performance ($Q{_\text{mech}}$)~\cite{thompson2008strong,tsaturyan2017ultracoherent,Richard2016,ghadimi2018elastic,shin2022spiderweb,bereyhi2022perimeter,Wilson2009,Cupertino2024}. As a result, Si$_3$N$_4$ membranes enable attonewton-level force sensing, chip-scale inertial navigation~\cite{krause2012high}, mechanical frequency combs~\cite{de2023mechanical}, and quantum state preparation at room temperature~\cite{mason2019continuous}. Their versatility has enabled a wide range of nanomechanical geometries, including: strings~\cite{verbridge2006high,sadeghi2019influence,schmid2011damping,shin2022spiderweb,bereyhi2022perimeter,hoch2022geometric,li2024strain,bereyhi2022hierarchical}, drumheads~\cite{serra2018silicon,Wilson2009,thompson2008strong,Villanueva2014,Chakram2014}, phononic-crystal (PtC)resonators~\cite{yu2014phononic,tsaturyan2017ultracoherent,reetz2019analysis,ghadimi2018elastic,Cupertino2024}, and trampoline membranes~\cite{Norte2015,Reinhardt2016,Richard2016,hoj2021ultra}. All examples exhibit record-low damping and strong isolation from room temperature thermomechanical noise.

A key ingredient behind this performance is the high-temperature Low Pressure Chemical Vapor Deposition (LPCVD) required to form stoichiometric Si$_3$N$_4$ with gigapascal tensile stress and parts-per-billion optical absorption~\cite{land2024sub}. These elevated temperatures achieve stoichiometry that suppresses absorption and yields a mechanical loss tangent among the lowest of any nanomechanical material in the kHz–MHz bending regime. While Si$_3$N$_4$ resonators already exhibit the highest $Q{_\text{mech}}$ at room temperature, their quality factor increases even further at cryogenic temperatures~\cite{yuan2015silicon,purdy2012cavity,chuang2004mechanical,seis2022ground,fischer2016optical}, reinforcing their role as a uniquely powerful platform for both quantum and precision sensing at all temperatures.  

Such extreme mechanical sensitivity, however, demands an equally stable optical interface for readout. Achieving sub-nanometer alignment between reflective Si$_3$N$_4$ membranes and conventional mirrors requires exceptional mechanical precision and thermal stability, typically enforced by extensive alignment infrastructure~\cite{zhou2023cavity,zhou2024ultrahigh,de2022coherent,enzian2023phononically,chen2017high,dumont2019flexure,bui2012high,mitra2024narrow}. As shown in FIG.~\ref{fig0}a-b, Existing implementations often rely on fabrication schemes that etch completely through the Si$_3$N$_4$ chip to open an optical access window~\cite{hyatt2025fabrication}, followed by manual alignment of the membrane with an external cavity structure. These conventional approaches are delicate, time-consuming, and difficult to scale or package.

Recent efforts to integrate high-performance nanomechanical mirrors above distributed Bragg reflectors (DBRs) have explored sacrificial-layer deposition strategies, as demonstrated in III–V systems~\cite{Kini2023}. However, III–V platforms remain limited by small wafer formats, CMOS incompatibility, and mechanical properties that degrade over time. Extending similar integration schemes to Si$_3$N$_4$ has been fundamentally constrained by material incompatibility: the high temperatures ($\approx$800–900~$^\circ$C) required for stoichiometric, high-stress Si$_3$N$_4$ growth exceed the thermal budget of conventional Ta$_2$O$_5$/SiO$_2$ DBRs, leading to film degradation and delamination. Attempts to directly bond Si$_3$N$_4$ membranes on mirrors have not maintained a high $Q_{mech}$ \cite{hornig2020monolithically,dumont2019flexure} and still required tedious alignment. These limitations have prevented the realization of a scalable, monolithic optomechanical cavity platform that maintains the exceptional optical and mechanical performance of Si$_3$N$_4$ membranes. Ultra-short cavity lengths have been long-sought for their ability to enhance optomechanical coupling in ways that depart from longer free-space cavities~\cite{mitra2024narrow,Kini2023} 

In this work, we overcome these long-standing challenges by introducing a monolithic cavity platform that directly integrates a suspended high-stress Si$_3$N$_4$ membrane above a thermally robust, low-loss LPCVD SiO$_2$/Si$_3$N$_4$ DBR. Using a defect-free amorphous-silicon sacrificial layer and a stiction-free dry plasma undercut, we achieve sub-micron cavity gaps between membrane reflector and DBR. These are self-aligned with nanoscale precision and are among the shortest free-space optomechanical cavities to date utilizing reflective Si$_3$N$_4$ membranes~\cite{mitra2024narrow,Kini2023,hornig2020monolithically}. The suspended membranes with about 1 gigapascal of tensile stress exhibit picometer-scale sag on the order of an atomic radius, naturally producing parallel optical interfaces and high fabrication yield. This establishes a scalable route to compact, high-coherence Si$_3$N$_4$ cavity optomechanics without the alignment and assembly bottlenecks of conventional membrane-in-the-middle systems.

\section{Design and Optical Optimization of an Integrated Optomechanical Cavity }

\begin{figure*}
\centering
\includegraphics[width=0.92\textwidth]{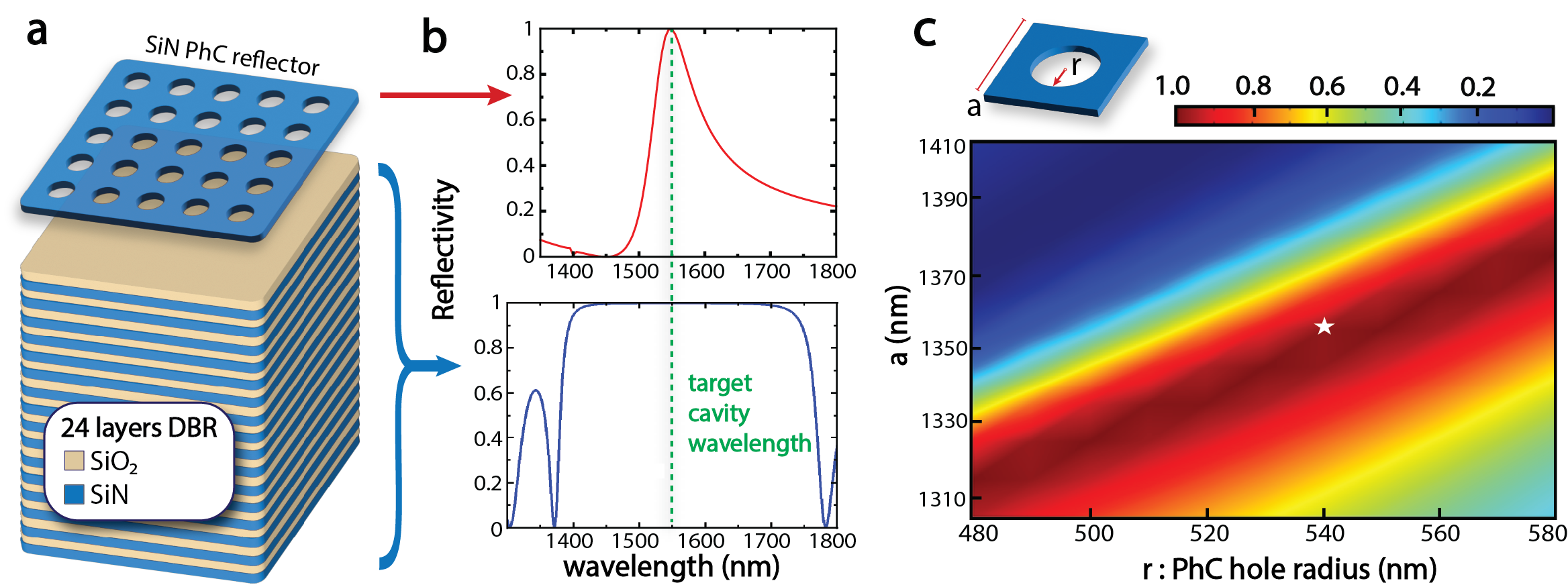}
\caption{Design and optical optimization of the integrated Si$_3$N$_4$–DBR cavity.
(a) Schematic of the cavity structure showing a high-stress Si$_3$N$_4$ PtC membrane suspended above a 24-layer Si$_3$N$_4$/SiO$_2$ distributed Bragg reflector (DBR).
(b) Simulated reflectivity spectra of the PtC membrane (red) and the DBR (blue), showing their overlap at the target cavity wavelength of 1550~nm (green dashed line).
(c) Simulated reflectivity contour map of the PtC as a function of lattice constant $a$ and hole radius $r$, indicating the design point (white star) used for fabrication.}
\label{fig1}
\end{figure*}

Optical cavity readout is achieved by vertically coupling a mechanically compliant Si$_3$N$_4$ PtC membrane to a high-reflectivity DBR, forming a compact Fabry–Pérot cavity (FIG.~\ref{fig1}a). In this geometry, the suspended PtC serves as the partially transmitting top mirror and the DBR as the highly reflective bottom mirror, while the sub-micron air gap between them defines the optical cavity. Small membrane displacements modulate the cavity resonance frequency, enabling sensitive interferometric readout. The corresponding optomechanical coupling strength, \(G=\partial\omega_c/\partial x\), increases as the optical mode volume is reduced, so tighter vertical confinement and smaller cavity gaps enhance displacement-to-frequency transduction.

First, the DBR is designed. The DBR is deposited across the full wafer and provides a broadband reflection band. We use alternating SiO$_2$ ($n_1 = 1.44$) and Si$_3$N$_4$ ($n_2 = 2.0$) quarter-wave layers centered at 1550 nm. Thicknesses of 269 nm (SiO$_2$) and 194~nm (Si$_3$N$_4$) yield constructive optical interference and a simulated peak reflectivity of $\approx 99.91\%$ for a 24-layer stack with total thickness $\approx$5.8 \si{\mu m}. The topmost layer is SiO$_2$, which improves selectivity during the SF$_6$ release step, minimizes surface roughness, and ensures that the DBR remains optically pristine throughout processing. A representative DBR spectrum is shown in FIG.~\ref{fig1}b.

The suspended membrane acts as a lithographically tunable photonic mirror: adjusting its PtC geometry allows fine control over its reflectivity and over the wavelength at which it overlaps the DBR stopband. We use a 200~nm-thick high-stress Si$_3$N$_4$ film, which balances reflectivity with mechanical dissipation (thinner membranes improve $Q_{mech}$ but reduce achievable optical reflectivity).

Finite-element simulations (COMSOL) were used to obtain the reflectivity of the PtC by sweeping hole radius $r$ and lattice constant $a$ near the 1550 nm design point. The resulting reflectivity map (Fig.~\ref{fig1}c) displays a narrow ridge of high reflectance corresponding to constructive out-of-plane interference. The geometry marked by the white star maximizes reflectivity while remaining robust against lithographic tolerances.

Because the PtC mirror relies on in-plane periodicity, it is naturally polarization dependent and its reflectivity varies with the incident beam’s divergence. Larger, less focused beams sample more periods with a narrower in-plane $k$-distribution, improving effective reflectance. In contrast, the DBR has only vertical periodicity and is polarization insensitive, providing uniform broadband reflectivity.

To model the combined cavity, rigorous coupled-wave analysis (RCWA) is used. In this simulation, realistic fabrication deviations from nominal values are included. Good agreement with simulations is obtained using the following representative parameters
a = 1.36 \si{\mu m},
$r = 0.557~\mu$m,
$t_{\text{SiN}} = 193$~nm,
$t_{\text{SiO}} = 271$~nm,
$n_{\text{SiN}} = 2.09$,
$n_{\text{SiO}} = 1.42$,
cavity spacing $t_{\text{cav}} \approx 0.95~\mu$m, and
membrane thickness $t_{\text{PtC}} = 191$~nm.
Small deviations arise naturally in LPCVD films but do not significantly impact finesse due to the broad DBR stopband and the gentle slope of the PtC reflectivity ridge.

Overall, the DBR provides a thermally robust, broadband mirror, while the lithographically tunable PtC allows precise spectral matching to the cavity wavelength. Together they form a monolithic, self-aligned cavity with strong vertical confinement suitable for high-sensitivity optomechanical readout. Vertical optomechanical coupling is particularly well matched to Si$_3$N$_4$, since its highest-$Q$ resonators rely on out-of-plane flexural modes set by ultrathin film thicknesses ($\approx$200 \si{nm}), easily realized by deposition but very difficult to achieve through lateral nanolithography.

\section{\label{sec-6}{Scalable Nanofabrication Using Dry Processing Techniques}}

\begin{figure*}[t]
\centering
\includegraphics[width=0.92\textwidth]{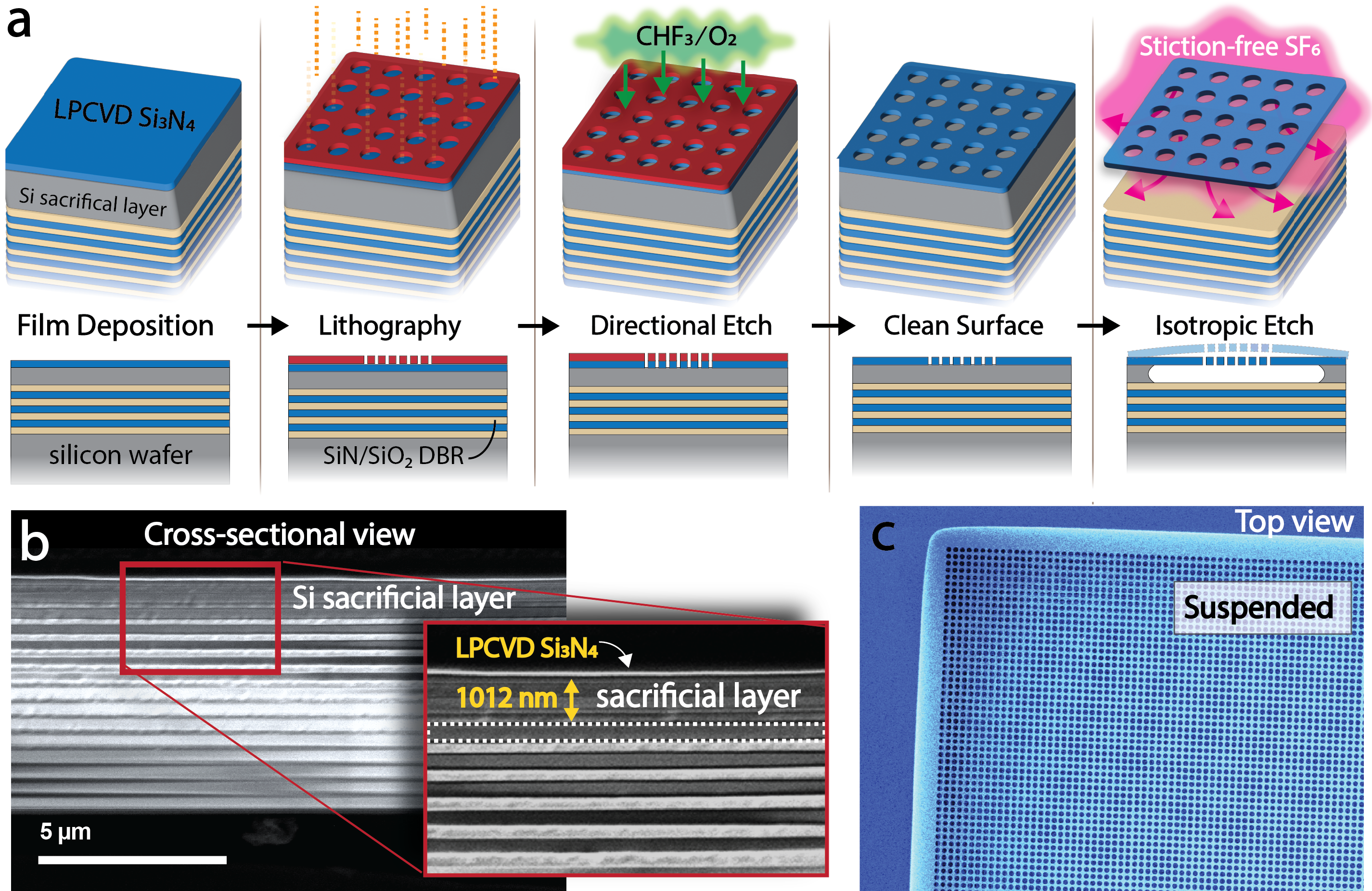}
\caption{\textbf{High-yield fabrication of monolithic Si$_3$N$_4$–DBR optomechanical cavities.}
(a) Schematic fabrication flow with top and side views for each step: LPCVD growth of the Si$_3$N$_4$/SiO$_2$ DBR and amorphous-Si sacrificial layer; lithographic patterning of the Si$_3$N$_4$ membrane; directional CHF$_3$/O$_2$ etching; surface cleaning; and stiction-free isotropic SF$_6$ undercut to suspend the membrane.
(b) Cross-sectional SEM showing the DBR stack, amorphous-Si sacrificial layer, and top Si$_3$N$_4$ membrane prior to release (inset: measured sacrificial-layer thickness $\sim$1012 \si{nm}). The white dashed lines highlight the top SiO$_2$ layer of the DBR.
(c) Top-view optical micrograph of the fully suspended membrane, demonstrating wafer-scale uniformity and clean release.}
\label{fig2}
\end{figure*}

To realize a robust, wafer-scale platform for monolithic optomechanical cavities, we developed a fabrication flow composed entirely of dry-processing steps (FIG. \ref{fig2}a), with both top- and cross-sectional views shown for each stage. Eliminating liquids in the fabrication process avoids stiction and capillary forces, ensuring that the pristine optical and mechanical properties of high-stress Si$_3$N$_4$ are fully preserved.

The process begins with LPCVD deposition of the multilayer Si$_3$N$_4$/SiO$_2$ DBR on a silicon wafer. Because every layer is deposited at high temperature, the DBR maintains its structural stability and sharp interfaces throughout subsequent processing. 1012 \si{nm} of an amorphous-Si sacrificial layer is then deposited by Inductively Coupled Plasma - Chemical Vapor Deposition (ICP-CVD), with parameters tuned to suppress hydrogen incorporation and prevent blistering or bubble formation during the later high-temperature LPCVD step. Finally, a 200 \si{nm} stoichiometric Si$_3$N$_4$ layer is deposited to form the high-stress membrane and later patterned with nanoholes to increase its reflectivity. The full deposition recipes are detailed in Methods.

The membrane pattern is defined using high-resolution electron-beam lithography. After exposure and development, the square-lattice PtC holes are transferred into the Si$_3$N$_4$ layer using a CHF$_3$/O$_2$ Inductively Coupled Plasma Reactive Ion Etching (ICP-RIE) that provides anisotropic profiles and smooth sidewalls. Residual resist is removed with N,N-dimethylformamide, followed by a piranha clean to eliminate organics and leave a chemically pristine membrane surface prior to suspension.

Suspension is achieved through a short isotropic SF$_6$ plasma undercut, which selectively removes the amorphous-Si layer while leaving both the membrane and DBR relatively unetched (FIG.~\ref{fig2}a). Because the release is fully dry, the process is inherently stiction-free, enabling the fabrication of high-aspect-ratio suspended nanostructures with minimal collapse or fracturing~\cite{Norder2025, shin2022spiderweb, Cupertino2024}. The SF$_6$ undercut produces exceptionally clean cavity interfaces, uniform air gaps, and high repeatability across full wafers~\cite{Guo2019}. Importantly, this dry-release method has repeatedly been shown to produce nanomechanical devices with some of the lowest mechanical and optical losses to date, and with high-fidelity agreement between simulated and measured dissipation—performance that is difficult to achieve with liquid-based etches such as KOH, which can require hours of through-wafer etching to obtain optical access. In contrast, our cavities are formed in tens of seconds, immediately yielding a fully aligned Si$_3$N$_4$ membrane--DBR cavity; this represents a fundamental shift in how state-of-the-art membrane materials are integrated with readily available optical readout systems.

Cross-sectional SEM images (FIG.~\ref{fig2}b) show the multilayer DBR, the sacrificial layer, and the sharpness of the interfaces produced by the LPCVD stack. Top-view images (FIG. \ref{fig2}c) show fully suspended membranes with uniform PtC patterns. The residual tensile stress of $\approx$1 GPa keeps the films taut, flat, and robust during handling and optical probing, with nanoscale sag across millimeter-scale spans~\cite{xu2025measuring}, producing cavity parallelism between the membrane and DBR that would be infeasible to recreate with conventional alignment schemes as shown in FIG.\ref{fig0}b.

Together, these dry-processing steps provide a high-yield, wafer-scale route to monolithic membrane--DBR cavities with robust structural integrity. We now show that the platform delivers more than mechanically surviving membranes: it also preserves strong optical cavity performance and low mechanical dissipation.

\begin{figure*}[t]
\centering
\includegraphics[width=1.0\textwidth]{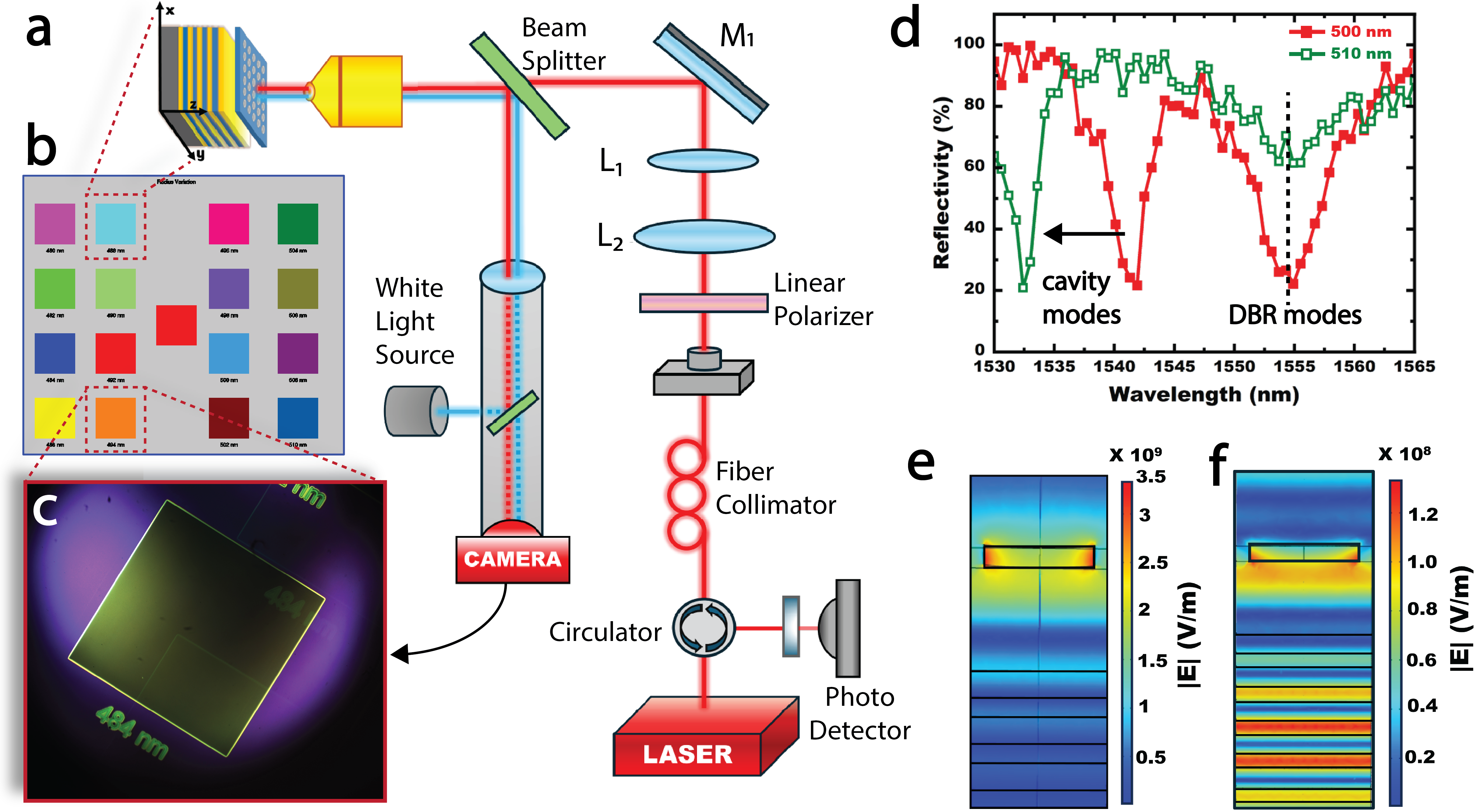}
\caption{Optical characterization of the integrated Si$_3$N$_4$–DBR cavity using broadband reflectivity measurements. (a) Schematic of the micro-reflectivity setup used for cavity measurements, combining a tunable IR laser for reflectivity spectroscopy with a white-light camera path for alignment.
(b) Layout of the fabricated chip showing an array of sixteen 1×1 mm$^2$ suspended PtC membranes with lithographically varied hole radii, plus a central DBR-only reference region.
(c) Camera image used during alignment, showing a suspended Si$_3$N$_4$ membrane under white-light illumination with the IR probe spot positioned at its center.
(d) Measured reflectivity spectra for two membrane designs (500 nm and 510 nm hole radii). The cavity mode shifts with radius, while the DBR-associated mode remains fixed.
(e) Simulated electric-field distribution of the vertically confined cavity mode formed in the air gap.
(f) Simulated electric-field distribution of the DBR-associated photonic band-edge mode, confined mainly within the patterned membrane.}
\label{fig3}
\end{figure*}

\section{\label{sec-7}{Characterization of Optical Cavity Resonances}}
To evaluate the optical performance of the integrated Si$_3$N$_4$~membrane–DBR cavity, we performed micro-reflectivity measurements using the setup shown in FIG.~\ref{fig3}a. A tunable diode laser (1530–1620 nm) is routed through a collimator, polarization controller, and polarizer, then focused onto the membrane using a high-NA objective. Reflected light returns through the same optical path and is routed to a photodiode via a circulator. A white-light illumination path and camera are integrated into the beam path, enabling real-time imaging and precise alignment of the membrane under test.

To allow systematic optical characterization, the chip contains an array of sixteen 1×1 mm$^2$ suspended membranes with lithographically varied PtC radii, as illustrated in FIG.~\ref{fig3}b. A reference region without a membrane (center tile) provides direct measurement of the bare DBR reflectivity. During measurement, the camera image (FIG.~\ref{fig3}c) is used to position the IR probe spot on the suspended membrane while visually confirming membrane integrity and avoiding edges or defects.

Representative reflectivity spectra for two different PtC radii are shown in FIG.~\ref{fig3}d. Each spectrum exhibits two prominent dips. The short-wavelength feature (1535–1545 nm) corresponds to the vertically confined cavity mode, and its position shifts noticeably when the hole radius is changed, reflecting the sensitivity of this mode to the membrane’s effective refractive index. In contrast, the longer-wavelength dip near 1555~nm originates from the DBR stack and remains spectrally fixed for all membrane designs, confirming that it is governed by the multilayer mirror rather than the patterned Si$_3$N$_4$ membrane.

The cavity mode, formed in the vertical air gap between the membrane and DBR, appears as the short-wavelength dip and shifts when the hole radius changes (500~nm vs. 510~nm). This shift reflects the change in the membrane’s effective refractive index as the air fill fraction increases. In contrast, the DBR mode is dictated by the multilayer stack and remains spectrally invariant.

To identify the physical origin of these resonances, we performed finite-element simulations of the electric-field distribution (FIG.~\ref{fig3}e–f). At the cavity resonance wavelength, the field is strongly localized within the air gap, forming a vertically confined Fabry–Pérot mode with a standing-wave pattern between the high-reflectivity DBR and the PtC membrane. This mode exhibits a narrow linewidth with measured FWHM of 2.55~nm, corresponding to an optical quality factor $Q_{\mathrm{opt}}\approx 604$ and a finesse of $\approx800$.

The longer-wavelength feature corresponds to a DBR mode that arises when light scattered by the patterned membrane acquires non-planar wavevectors. Although a bare DBR strongly reflects normally incident plane waves, the proximity of the PtC membrane introduces lateral scattering that generates off-axis and higher-order $k$-components. These components can penetrate the multilayer stack and form a band-edge–like optical mode primarily confined within the DBR, with only weak, evanescent coupling back into the membrane. Because this resonance is governed by in-plane scattering dynamics rather than the vertical cavity spacing, its spectral response remains fixed as the hole radius is varied. As a result, it appears as a broader, geometry-insensitive feature (FWHM 6.13~nm, $Q_{\mathrm{opt}}\approx254$) in the reflectivity spectrum.

Finally, FIG.~\ref{fig3}d demonstrates spectral tunability across devices on the same chip: increasing the PtC hole radius results in a blue-shift of the cavity mode while leaving the DBR mode unchanged. This confirms that the membrane governs the cavity-mode position, while the DBR provides a stable, broadband bottom mirror unaffected by lithographic variations.

Overall, the combined experimental and simulated results confirm that the integrated platform functions as a stable, tunable, high-finesse cavity. The cavity mode’s strong out-of-plane localization makes it ideal for interferometric displacement sensing, while the coexistence of membrane and DBR modes offers opportunities for hybrid photonic–optomechanical architectures.

\section{Mechanical Resonance Characterization and High-Q Performance of Suspended Membrane Devices}

 In the past several years, the isotropic SF$_6$ plasma undercut used here has emerged as one of the most powerful release techniques for Si$_3$N$_4$, enabling devices with $Q_{\mathrm{mech}}$ approaching $10^9$–$10^{10}$ on standard silicon substrates. However, it has never been established whether integrating Si$_3$N$_4$ with a multilayer DBR—containing dozens of interfaces, different acoustic impedances, and altered boundary conditions—introduces mechanical loss or degrades dissipation dilution. Demonstrating Q preservation is therefore essential before this integration platform can be used for state-of-the-art optomechanically engineered geometries.

\begin{figure*}[t]
\centering
\includegraphics[width=1\textwidth]{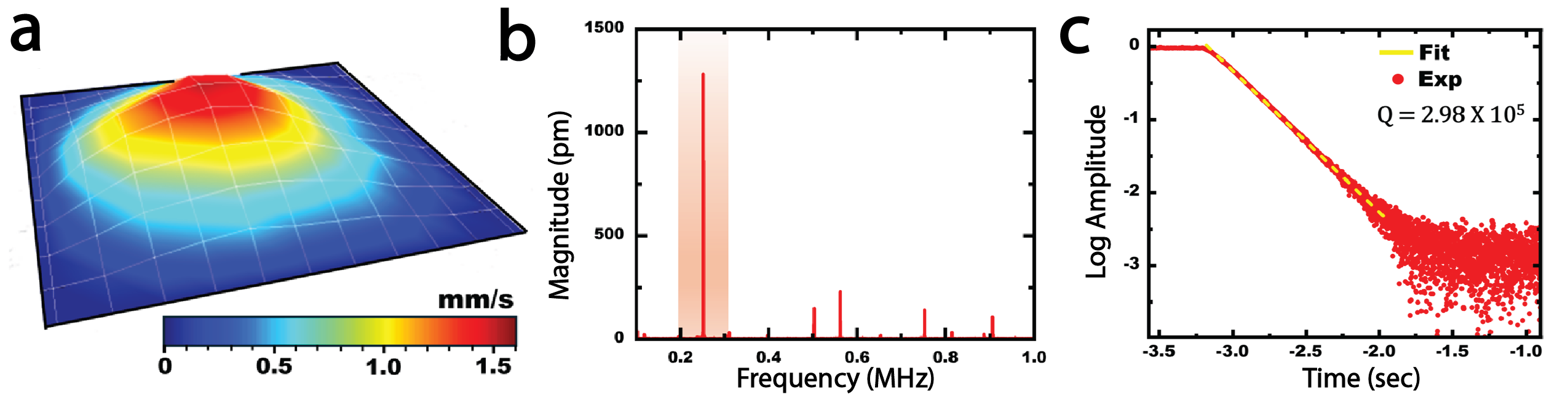}
\caption{Mechanical characterization of the suspended Si$_3$N$_4$ membrane integrated with a DBR platform.  
(a) Out-of-plane velocity map of the fundamental membrane mode measured by laser Doppler vibrometer.  
(b) Frequency-domain spectrum showing the fundamental resonance and higher-order modes.  
(c) Mechanical ringdown measurement of the fundamental mode, with an exponential fit (yellow) used to extract $Q_{\mathrm{mech}}$.}
\label{Figure4}
\end{figure*}

To evaluate this, we characterized DBR-integrated membranes using laser Doppler vibrometry (LDV) under high vacuum ($\approx10^{-6}$ mbar). FIG.~\ref{Figure4}a shows the measured out-of-plane velocity field of the fundamental membrane mode, exhibiting the expected symmetric displacement profile. A broadband sweep (FIG.~\ref{Figure4}b) identifies the fundamental resonance near 259 kHz and several higher-order modes extending to 1 MHz. The fundamental peak was then isolated, and its decay was measured after abruptly switching off the piezo drive. The ringdown trace (FIG.~\ref{Figure4}c) yields a mechanical quality factor $Q_{\mathrm{mech}} = 3.0 \times 10^{5}$, demonstrating low intrinsic damping and strong stress dilution.

Because the membranes used here are 200~nm thick (i.e. chosen to ensure high optical reflectivity), these $Q_{mech}$ values are naturally lower than those achievable with ultra-thin (20 nm) Si$_3$N$_4$ films. Thicker mirrors are required for the PtC reflectivity, but multi-thickness designs can decouple these requirements (e.g., thick mirror region with thin, stress-diluted clamping)~\cite{Richard2016,huang2024room}. Thus, it is important not only to measure the $Q_{mech}$ of this simple square membrane, but to verify that DBR integration does not introduce additional loss. If $Q_{mech}$ is preserved in this geometry, the platform can support the full family of high-$Q_{mech}$, phononic-engineered, soft-clamped, or tethered designs known to reach $10^8$–$10^{10}$.

To assess the mechanical impact of DBR integration, we first identified the highest-$Q_{mech}$ device on the DBR-integrated chip, corresponding to a hole radius of 484~nm. We then fabricated the nominally identical chip layout without DBR integration and directly compared the same membrane design (FIG.~\ref{fig3}b). The DBR-integrated membrane exhibited a fundamental frequency of 259~kHz and $Q_{mech} = 3.0 \times 10^5$, while the corresponding membrane on silicon without DBR showed a slightly higher frequency of 290~kHz and $Q_{mech} = 3.8 \times 10^5$. The frequency shift is consistent with small variations in film thickness and thermal-expansion stresses associated with the DBR structure. Across the measured device sets, the average mechanical quality factors were $\langle Q_{mech} \rangle = 1.3 \times 10^5$ for the PhC+DBR devices and $\langle Q_{mech} \rangle = 4.6 \times 10^5$ for the PhC-only devices, indicating that the integrated structures remain within the same order of magnitude in mechanical coherence. Crucially, no catastrophic degradation in $Q_{mech}$ was observed upon DBR integration. Unlike soft-clamped modes, the fundamental modes of square membranes are strongly coupled to the substrate and therefore maximize chip-based loss channels, making them a stringent reference for assessing additional dissipation introduced by the DBR stack.~\cite{de2022mechanical}

These measurements demonstrate that the Si$_3$N$_4$/SiO$_2$ DBR introduces little additional mechanical loss. The SF$_6$ release process preserves the dissipation characteristics of the membrane even on a multilayer mirror substrate, confirming full compatibility with high-stress Si$_3$N$_4$. This establishes the platform as a mechanically benign integration scheme capable of supporting the highest-Q geometries developed in the community while simultaneously enabling a compact, high-finesse optical cavity.

\section{Conclusion and Outlook}

We have presented a scalable, CMOS-compatible fabrication strategy that monolithically integrates high-stress Si$_3$N$_4$ membranes with broadband distributed Bragg reflectors (DBRs) using a wafer-level, lithography-defined process. This platform removes a longstanding bottleneck in cavity optomechanics by eliminating bonding, through-wafer etching, and precision alignment. Instead, a stiction-free gas-phase undercut defines a vertically self-aligned cavity with uniform sub-micron membrane--DBR spacing. Vertical optomechanical coupling is especially well suited to Si$_3$N$_4$, as its highest-$Q_{mech}$ resonators exploit out-of-plane flexural modes defined by ultrathin film thicknesses ($\approx$20~nm), which are straightforward to realize by deposition but extremely challenging to achieve through lateral nanolithography~\cite{Guo2019}.

A central materials advance is the use of alternating LPCVD Si$_3$N$_4$ and SiO$_2$ layers to form a thermally robust DBR. Although the refractive-index contrast is lower than in Ta$_2$O$_5$/SiO$_2$ stacks, the LPCVD process yields exceptionally low optical absorption and remains stable through the 800--900~°C required for stoichiometric high-stress Si$_3$N$_4$ deposition. In this way, lower index contrast is offset by low optical loss, enabling finesse values comparable to those demonstrated in higher-contrast III--V systems~\cite{Kini2023}.

Optical and mechanical characterization confirms that integration preserves the essential performance of the Si$_3$N$_4$ platform. Mechanical ringdown measurements show $Q{_\mathrm{mech}}>10^{5}$, comparable to non-integrated membranes, while reflectivity spectra show cavity finesses above 800 and strong resonances suitable for precision optical readout. Although our demonstration uses relatively thick (200~nm) square membranes rather than geometries optimized for record $Q_{mech}$, the DBR process adds little mechanical loss compared to identical devices without DBR.

The fabrication route is broadly compatible with the wider ecosystem of Si$_3$N$_4$ resonators used in cavity optomechanics, including reflective phononic membranes~\cite{enzian2023phononically}, trampolines~\cite{Richard2016}, hierarchical metamaterials, and high-aspect-ratio resonators~\cite{Norder2025}. It is also well suited to thick-substrate implementations, which are known to improve $Q_{mech}$ for modes without soft clamping \cite{Richard2016,wilson2012cavity,de2022mechanical}, since the SF$_6$ gas-phase release suspends membranes on millimeter-scale substrates in seconds while avoiding the stiction, bowing, and poor yield associated with wet-etch access through thick chips~\cite{Norte2015}. Previous strategies based on locally thickening the mirror region while thinning the clamping region to tens of nanometers~\cite{Richard2016,huang2024room} should therefore be naturally extendable to this platform.

More broadly, the platform is compatible with the ultra-high-aspect-ratio geometries enabled by high-stress Si$_3$N$_4$. Recent work has shown that centimeter-scale PtC reflectors with nanometer thickness and billions of holes can be fabricated with exceptional flatness~\cite{Norder2025}; at these scales, larger beam diameters can sustain high reflectivity even for thinner membranes~\cite{moura2018centimeter}. Although we demonstrate a flat--flat cavity here, PtC structures with radially varying hole geometries can also impart effective curvature~\cite{agrawal2024focusing,guo2017integrated}, offering future opportunities for transverse-mode engineering, reduced sensitivity to lateral misalignment, and lower clipping loss.

Taken together, this monolithic platform transforms a delicate, alignment-limited cavity optomechanical architecture into a robust and manufacturable one. By preserving the strong optical and mechanical performance of high-stress Si$_3$N$_4$ while replacing complex assembly with a simple self-aligned process, it provides a practical route toward scalable precision sensing, integrated photonics, and next-generation quantum technologies.

\section{Methods}
The DBR multilayer was fabricated by alternating low-pressure chemical vapor deposition (LPCVD) of SiO$_2$ and low-stress silicon nitride layers. The oxide films were deposited in a horizontal hot-wall LPCVD furnace at 700~$^\circ$C and a pressure of 150~mTorr, using vaporized tetraethyl orthosilicate (TEOS) as the precursor. Low-stress silicon nitride layers were deposited in a similar furnace by introducing a mixture of NH$_3$ and dichlorosilane (DCS), with a fixed NH$_3$/DCS flow ratio of 0.19, corresponding to a Si-rich regime. The nitride deposition was carried out at 860~$^\circ$C and 150~mTorr.

LPCVD was preferred over plasma-enhanced CVD (PECVD) for nitride deposition due to the negligible hydrogen incorporation associated with the former technique. The resulting low hydrogen content confers excellent thermal stability to the films, preventing significant stress variations as well as cracking or delamination during subsequent deposition steps.

After completion of the DBR stack, a 1~$\mu$m-thick amorphous silicon layer was deposited in an LPCVD reactor by flowing SiH$_4$ at 560~$^\circ$C. The topmost high-stress silicon nitride layer was subsequently deposited in the same furnace used for the DBR nitride layers, but with a substantially higher NH$_3$/DCS flow ratio corresponding to stoichiometric conditions. In this case, the deposition temperature and pressure were set to 700~$^\circ$C and 150~mTorr, respectively.

\section*{Acknowledgments}

We thank Mohammadjavad Ebrahimipour, Juan Lizarraga Lallana for experimental help and  Gerard Verbiest, Peter Steeneken, Andrea Cupertino and Dalziel Wilson for thoughtful comments. This work has received funding from the EMPIR programme co-financed by the Participating States and from the European Union’s Horizon 2020 research and innovation programme (No. 17FUN05 PhotoQuant). This publication is part of the project, Probing the physics of exotic superconductors with microchip Casimir experiments (740.018.020) of the research programme NWO Start-up which is partly financed by the Dutch Research Council (NWO). Funded/Co-funded by the European Union (ERC, EARS, 101042855). Views and opinions expressed are however those of the author(s) only and do not necessarily reflect those of the European Union or the European Research Council. Neither the European Union nor the granting authority can be held responsible for them.

\section*{Conflict of Interest}
The authors declare no conflict of interest.

\section*{Data Availability Statement}
The data that support the findings of this study are available from the corresponding author upon reasonable request.

\appendix

\bibliography{apssamp}

\appendix

\onecolumngrid

\newpage

\end{document}